# Numerical Modeling of a Teeth-shaped Nano-plasmonic Waveguide Filter

## Xianshi LIN and Xuguang HUANG[*]

*Laboratory of Photonic Information Technology, South China Normal University, Guangzhou, 510006, China*

[*]*Corresponding author: huangxg@scnu.edu.cn*

In this paper, tooth-shaped and multiple-teeth-shaped plasmonic filters in the metal-insulator-metal (MIM) waveguides are demonstrated numerically. By introducing a three-port waveguide splitter, a modified model based on the multiple-beam-interference and the scattering matrix is given. The transmittance spectrum as a function of teeth width, depth, period and period number are respectively addressed. The result shows the new structure not only performs the filtering function as well as MIM grating-like structures, but also is of submicrometer size for ultra-high integration and relatively easy fabrication. © 2009 Optical Society of America

## 1. Introduction

Plasmons are highly localized to a metal surface that can have applications as sub-wavelength waveguides to guide light below the diffraction limit in conventional optics [1-3]. Thus, plasmonic waveguides based on metallic nano-structures have shown the potential to increase device densities in integrated photonic circuits, which may one day replace the current electronic integrated circuits. Many interesting experimental or theoretical works on SPPs waveguiding structures have been reported so far, such as metallic stripes and wires [4], grooves in a metal surface [5] and a chain of metallic nanoparticles [6]. Waveguides consisting of an insulator sandwiched between two metals serve as metal-insulator-metal (MIM) waveguides support propagating surface plasmon modes that are strongly confined in the insulator region with an acceptable propagation length [7]. Taking advantage of MIM waveguides, a variety of functional plasmonic structures can be designed and fabricated right now, such as U-shaped waveguides [8], splitters [9], Y-shaped combiners [10], couplers [11] and M-Z interferometers [12,13].

Recently, the interest in the resonators [14], metal heterostructure structures [15-17] and Bragg reflectors [18-23] formed in MIM waveguides to achieve wavelength filtering function has been further fueled by the theoretical predictions and experimental demonstrations. Given the perspective of integrating various functional components within several micrometers, people need to decrease the size of a device to meet the demand of high integration. However, most of the filter structures mentioned above have large sizes over several wavelengths with the period number of N > 9 for the grating-like structures. Large size or length also results in relatively high propagation loss. In order to solve these problems, stub structures based on MIM waveguides which can function as wavelength selective filters of a submicron size have been numerically presented [24]. More recently, we proposed and demonstrated numerically a new nanometeric plasmonic filter in a single tooth-shaped MIM waveguide [25] to achieve an ultracompact size with hundreds of nanometers in length and low insertion loss. Only a simply analytic model based on multiple-beam-interference and the scattering matrix method is given without further discuss, due to the limitation of length.

In this paper, we modify the model of the tooth-shaped structure by taking into account wavelength dependences of all of the reflection, transmission, and splitting coefficients and their propagation losses of each port of the structure, and analyze the response profile of the structure as wavelength in comparing with the transmittance spectrum resulted from the simulation in Finite-Difference Time-Domain (FDTD) method. We further extend our previous work to multiple-teeth-shaped structures to achieve better filter response.

In section 2, the effective index of a symmetric MIM waveguide is numerically analyzed. In section 3, a modified model based on the scattering matrix is given by introducing a three-port waveguide splitter, and the filtering characteristics of the tooth-shaped waveguide filter are investigated to compare with the simulation in FDTD method. In section 4, transmittances with wide bandgap around 1.55μm of multiple-teeth-shaped MIM waveguide filters are simulated. The transmittance spectrum as a function of teeth width, depth, period and period number are respectively discussed.

## 2. Characterizes of a MIM waveguide

Firstly, we consider the SPPs traveling in a two-dimensional MIM slit waveguide along the z-axis direction shown in the inset of Fig. 1(a). Only the TM mode consisting of $E_x$, $E_z$ and $H_y$ components is discussed due to its obvious plasmon excitation on the metallic surfaces. The dispersion equation for TM mode in the waveguide is given by [18,26]

$$\varepsilon_d k_{z2} + \varepsilon_m k_{z1} \coth(-\frac{ik_{z1}}{2}w) = 0 \tag{1}$$

with $k_{z1}$ and $k_{z2}$ defined as: $k_{z1}^2 = \varepsilon_d k_0^2 - \beta^2$ and $k_{z2}^2 = \varepsilon_m k_0^2 - \beta^2$, where $\varepsilon_d$ and $\varepsilon_m$ are respectively dielectric constants of the dielectric media and the metal, respectively. $k_0 = 2\pi / \lambda_0$ is free-space wave vector. The propagation constant $\beta$ is represented as effective index $n_{eff} = \beta / k_0$ of the waveguide for SPPs. In the paper, the dielectric is assumed to be air with $\varepsilon_d = 1$, and the metal to be silver. The dielectric constants $\varepsilon_m$ of silver can be calculated by Drude model [23]:

$$\varepsilon_m(\omega) = \varepsilon_\infty - \frac{\omega_p^2}{\omega(\omega + i\gamma)}. \tag{2}$$

Here $\varepsilon_\infty$ stands for the dielectric constant at infinite angular frequency with the value of 3.7, $\omega_p = 1.38 \times 10^{16} Hz$ is the bulk plasma frequency, which represents the natural frequency of the oscillations of free conduction electrons. And $\gamma = 2.73 \times 10^{13} Hz$ is the damping frequency of the oscillations, $\omega$ is the angular frequency of the incident electromagnetic radiation. It should be noted that the Drude model is not accurate enough for sliver and cause error, especially for infrared wavelengths. The more appropriate experimental data could be found in the handbook [27].

To fully understand how the width of waveguide influences the SPPs propagation, the dependence of the effective index of SPPs of a slit waveguide on the wavelength of the incident light with various width of the slit is presented in Fig. 1. Figure 1(a) shows the real part of the effective indexes $n_{eff}$ of the slit waveguide as a function of the wavelength at different widths of the slit. It can be seen that the real part decreases rapidly with the increase of wavelength from 400nm to 800nm, and then becomes saturation. The imaginary part of $n_{eff}$ is referred to the propagation length which is defined as the length over which the power carried by the wave decays to 1/e of its initial value: $L_{spps} = \lambda_0 / \left[ 4\pi \text{Im}(n_{eff}) \right]$, as shown in Fig. 1(b). It is clearly that smaller slit will have higher loss and shorter propagation length, although the field confinement of the slit waveguide is stronger.

## 3. A nanoscale tooth-shaped plasmonic waveguide filter

Figure 2(a) shows the typical structure of a proposed single tooth-shaped waveguide filter. The transmittance is defined to be $T=P_{out}/P_{in}$, where $P_{out}$ is the output power of a tooth-shaped structure or any other structure and $P_{in}$ is the input power of a straight MIM waveguide with the same length as the filter. $P_{out}$ and $P_{in}$ are calculated by integrating the normal component of the Poynting vector along the dashed lines. With the definition, the effect of the coupling between light source and MIM waveguide can be excluded. In the FDTD simulations, the grid sizes in the x and z directions are chosen to be 5nm×5nm. The length of $L$ is fixed by 300nm and the tooth width $w_t$=50nm. The transmission characteristics of the filter is investigated by choosing $d$=100nm and $d$=300nm respectively. As we can see from the Fig. 2(b), for the filter with $d$=100nm, the though of the transmittance is at the wavelength of 784nm. Two troughs with the transmittance ~0% occur at the wavelengths of 682nm and 1922nm for the filter with $d$=300nm, while the maximum transmittance at the wavelengths around 900nm is over 95%.

A simplified analytic model to explain the filtering function of the structure based on multiple-beam-interference and the scattering matrix [28] has been established in our earlier publication [25], the transmittance $T$ from Port 1 to Port 2 is derived as following:

$$T = \left| t_1 + \frac{s_1 s_3}{1 - r_3 \exp(i\phi(\lambda))} \exp(i\phi(\lambda)) \right|^2 \qquad (3)$$

where $r_i$, $t_i$ and $s_i$ ($i$=1,2,3) are respectively the reflection, transmission and splitting coefficients of a incident beam from Port $i$ ($i$=1,2,3) [see Fig. 2(a)]. The phase delay in the Port 3 of $\phi(\lambda)$ is modified as following:

$$\phi(\lambda) = 2\pi \cdot (n_{eff} \cdot 2d + w) / \lambda + \Delta\phi \qquad (4)$$

$\Delta\phi$ is an additional phase shift with wavelength-dependence which is produced in the propagation. The wavelength $\lambda_m$ of the $m$-th minimum of the transmission is given by:

$$\lambda_m = \frac{2(n_{eff} \cdot 2d + w)}{(2m+1) - \dfrac{\Delta\phi}{\pi}}, \quad (m = 0,1,2,...) \tag{5}$$

as the phase of $\phi(\lambda)$ at the wavelength of $\lambda_m$ equals to $(2m+1)\pi$. In our earlier publication, all of the scattering coefficients do not include any absorption of the MIM waveguide. And they satisfy the relation: $|r_i|^2 + |s_i|^2 + |t_i|^2 = 1$ in this lossless situation. Additionally, the simple model dose not taken into account any dispersion effect of scattering coefficients $r_i$, $t_i$ and $s_i$, as well as propagation loss. However, all of those parameters are wavelength dependent in real application.

To improve/modified the model of the filter, all of the parameters need to be carefully solved under the existences of both absorption loss and wavelength dispersion of MIM waveguides. To obtain accurate values of $\{r_i, t_i, s_i\}$ of the three ports, the numerical calculation based on dispersive and absorptive FDTD method is employed as follows. Port 3 is replaced by connecting a side-branch onto the main waveguides as a three-port waveguide splitter in dispersive and absorptive Ag metal material (shown inset of Fig. 3). The widths of the waveguide and side-branch waveguide are chosen to have the same value $w$=50nm. Figure 3(a) shows the normalized transmittance $T_{Port1-Port2} = |t_1|^2 = P_2^{out} / P_{in}$ and $T_{Port1-Port3} = |s_1|^2 = P_3^{out} / P_{in}$ of Port 2 and Port 3 at various wavelengths while only having input into Port 1. The result shows that output power of Port 2 is increase fast for wavelength range 400nm to 600nm and becomes finally saturated after 800nm and larger than the output power from Port 3 all along. As showing in Fig. 3(b), when only having input into Port 3, the normalized transmittance of Port 1 and Port 2 are equal and given as $T_{Port3-Port1} = T_{Port3-Port2} = |s_3|^2 = P_1^{out} / P_{in}$, due to the symmetry of the splitter. Due to the absorptive character of the metal, the calculation of reflection coefficient $r_i$ should consider the modified factors of propagation losses of { $\exp(-\alpha l_t)$, $\exp(-\alpha l_s)$, $\exp(-\alpha l_r)$ } for the transmitted, split, and reflected beams, and is given by $|r_i|^2 = [1 - |s_i|^2 / \exp(-\alpha l_s) - |t_i|^2 / \exp(-\alpha l_t)]\exp(-\alpha l_r)$. Here, $\alpha$ is the factor of propagation loss of MIM waveguide, $l_t, l_s$ and $l_r$ are respectively the propagation distances of the transmitted, split, and reflected beams from their input port to output port.

Figure 4 shows the transmittances of the tooth-shaped waveguide filters with two different tooth depths, calculated from the modified semi-analytic model, comparing with the FDTD results and the simple model. It can be seen that the results of the modified semi-analytic model agree with the FDTD simulations, while there is significant deviation of the simple model from the FDTD simulations, especially for the positions of the minimums. It means that the filtering mechanism and characteristics of the tooth-shaped can be well described with this modified model.

## 4. A multiple-teeth-shaped waveguide filter

It is straight forward and basic interest to expand a single tooth structure to multiple-teeth structure (shown in Fig. 5(a)), and check the difference between them. For the sake of comparison, the waveguide width $w$ and the distance $L$ are fixed to be 50nm and 300nm. $\Lambda$ and $N$ are the period and the number of rectangular teeth, respectively. $w_{gap}$ stands for the width of the gap that between any two adjacent teeth, and one has $w_t + w_{gap} = \Lambda$. A typical transmittance of the multiple-teeth-shaped waveguide filter with $w_t$=50nm, $\Lambda$=150nm, $d$=260.5nm and $N$=5 is shown in the Fig. 5(b), which is obtained with FDTD method. A wide bandgap occurs around $\lambda$ =1.55μm with the bandgap width (defined as the difference between the two wavelengths at each of which the transmittance is equal to 1%) is 590nm, and the transmittance of passband is over 90%. The filter's feature can be attributed to the superposition of the reflected and transmitted fields from each of the five single tooth-shaped components. Figure 6(a) shows the central wavelength of the transmittance bandgap, while the right y-axis displays the bandgap width of the waveguide filter as a function of teeth depth $d$ at various $w_t$ with the same $\Lambda$=150nm and $N$=5. The FDTD simulation results reveal that the relationship between the central wavelength of the bandgap and the teeth depth $d$ is a linear function for any $w_t$, which is indeed the one of expectations in Eq. (5). Figure 6(b) shows the central wavelength of the bandgap and the bandgap width as a function of teeth width of $w_t$ at various teeth depths. As revealed in the Eq. (5), the relationship between the bandgap position and $w_t$ mainly results from the contribution of the inverse-proportion-like dependence of $n_{eff}$ on $w_t$ as shown in Fig. 1(a). Obviously, teeth width $w_t$ should be chosen within the range of 50-100nm with a slope of $d\lambda_{center} / dw_t \approx 5$ to avoid a large value of $d\lambda_{center} / dw_t \approx 20$ when $w_t$<45nm in Fig. 6(b), and to reduce the sensitivity of the central wavelength of bandgap in fabrication process. Therefore, one can realize the filter function at various required wavelengths with high performance, by choosing the width or/and the depth of the teeth.

For the multiple-teeth-shaped structure with the parameters of $n_{eff,teeth}$ =1.070 for the width of $D$=$d$+$w$=(260.5+50)nm, $n_{eff,wg}$ =1.375 for $w$=50nm, and the teeth period of $\Lambda = w_t + w_{gap}$ =150nm in z-axis direction, one can see $w_t \, \mathrm{Re}(n_{eff,teeth}) + w_{gap} \, \mathrm{Re}(n_{eff,wg}) \approx$ 401.1nm $< 1550$nm$/2$. Thus the structure does not follow the Bragg condition in z-axis direction. However, the Bragg condition holds in each of teeth along x-axis direction as shown in Eq. (5). The function of small number of teeth is just to enhance and optimize filter's feature.

Figure 7(a) and (b) show the spectra of the transmittance of a multiple-teeth-shaped waveguide filter at different periods $\Lambda$ and period numbers $N$. As one can see from Fig. 7(a), when $\Lambda$=100nm is chosen, the coupling of the SPPs waves between two adjacent teeth is strong which causes the central bandgap wavelength to shift left and the bandgap to be wider. When the period equals to $\Lambda$=200nm, the coupling between any two adjacent teeth becomes very weak. One can see in Fig. 7(b) that, the forbidden bandwidth increases little with the changing of the period number from $N$=3 to 7, while the transmittance of the passband decreases from 93% to 86%. The reason for the decreasing in transmittance can be attributed to the increasing of the propagation loss of the lengthened structure with a large period number. From the simulation results, a tradeoff period number $N$=4 is the optimized number with the transverse filter length of 4×150nm, which is ~5 times shorter than the previous grating-like filter structures.

## 5. Conclusion

In this paper, tooth-shaped and multiple-teeth-shaped plasmonic filters in the MIM waveguides are demonstrated numerically. A modified model based on the multiple-beam-interference and the scattering matrix is given, and the transmittance response of the structure as wavelength is analyzed in comparing with the simulated transmittance spectrum. The transmittance spectrum of the multiple-teeth-shaped filter as a function of teeth width, depth, period and period number are all addressed respectively. It is possible to use the new periodic structure to achieve plasmonic wavelength filters with extreme high integration for the integrated optical circuits.

## Acknowledgment

The authors acknowledge the financial support from the Natural Science Foundation of Guangdong Province, China (Grant No. 07117866).

**Figure Captions**

Fig. 1.  (a) Real part of the effective index of refraction versus the width of a MIM slit waveguide structure. (b) Propagation length as a function of wavelength with different width of a MIM slit waveguide structure.

Fig. 2.  (a) The structure schematics of a single tooth-shaped waveguide filter, with the slit width of $w$, the tooth width of $w_t$, and the tooth depth of $d$. (b) The transmittance of the filter with $d$=100nm and $d$=300nm at a fixed $w$=50nm, $w_t$=50nm.

Fig. 3. The normalized transmittance of a three-port waveguide splitter with $w$=50nm by having input into Port 1 for (a) and Port 3 for (b).

Fig. 4. The comparison of the modified semi-analytical model with the FDTD results and the simple model.

Fig. 5. (a) Schematic of a multiple-teeth-shaped MIM waveguide structure. (b) The transmittance of the multiple-teeth-shaped waveguide filter with $w_t$=50nm, $\Lambda$=150nm, $d$=260.5nm and $N$=5.

Fig. 6.  (a) The central wavelength of the bandgap and the bandgap width as a function of the teeth depth of $d$ at various teeth widths. (b) The central wavelength of the bandgap and the bandgap width as a function of teeth widths of $w_t$ at various teeth depths.

Fig. 7. (a) Transmittance spectra of multi-teeth filters with different periods and a fixed *N*=5, (b)

Transmittance spectra of multi-teeth filters consisting of 3-7 periods with a fixed *Λ*=150nm.

**Fig. 1**

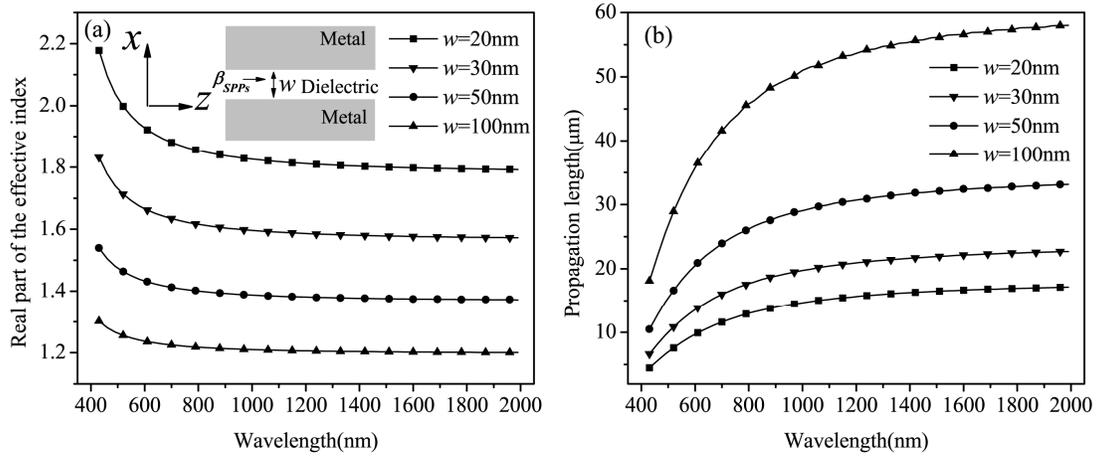

**Fig. 2**

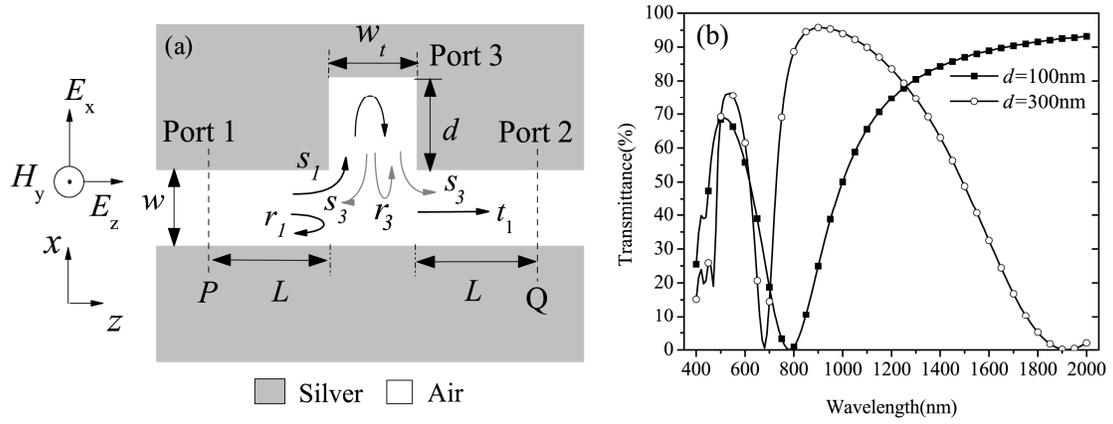

**Fig. 3**

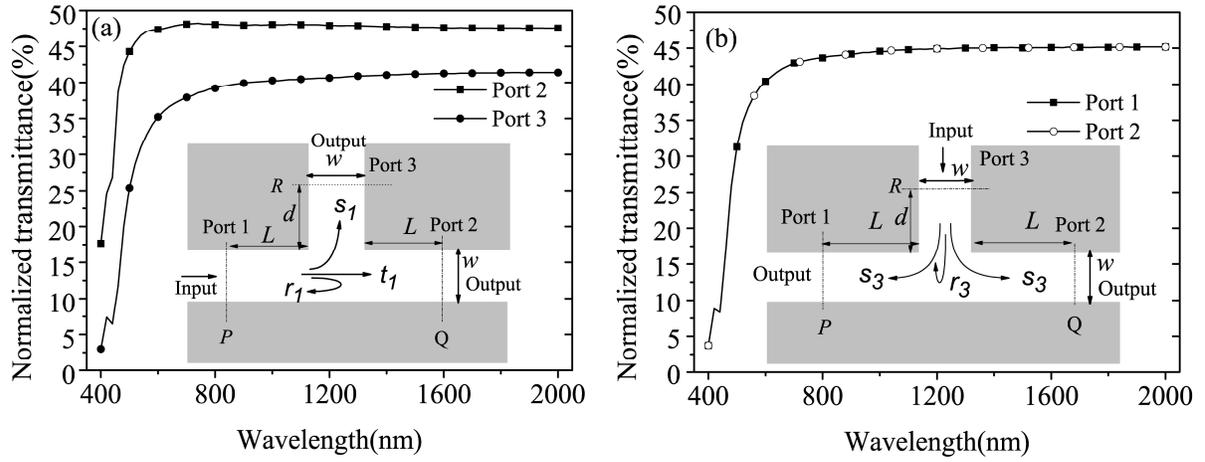

**Fig. 4**

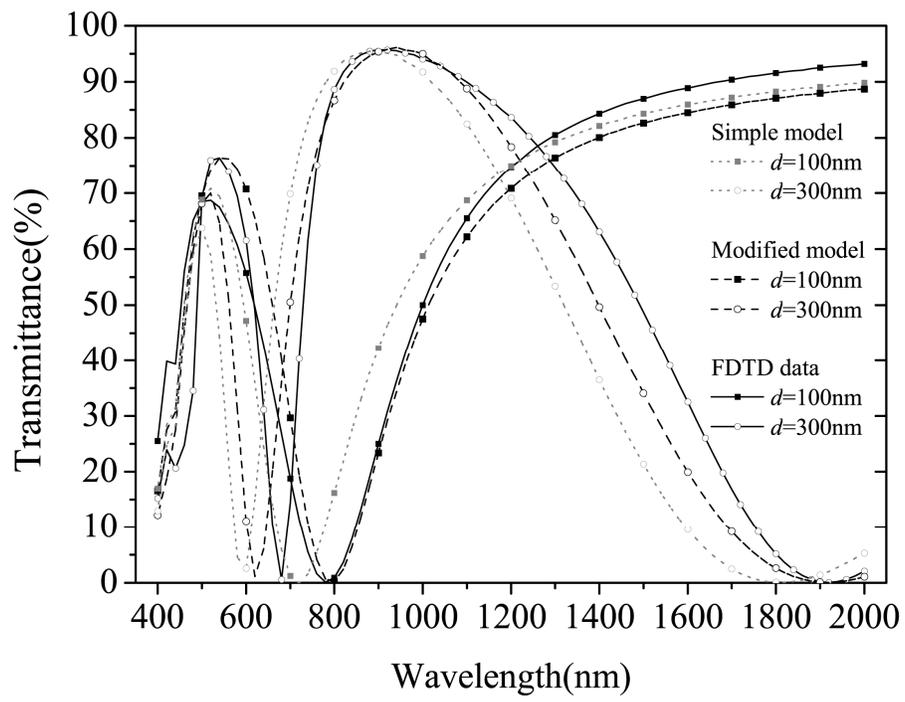

**Fig. 5**

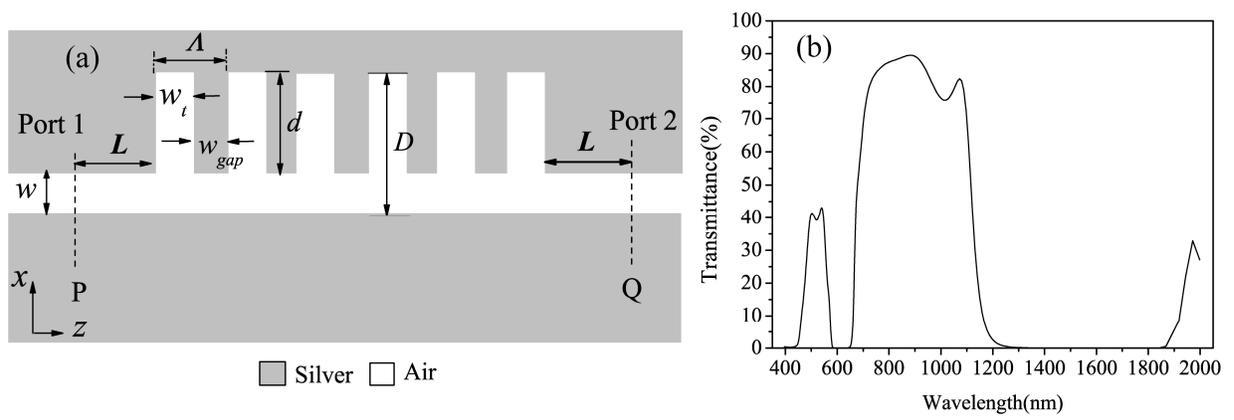

**Fig. 6**

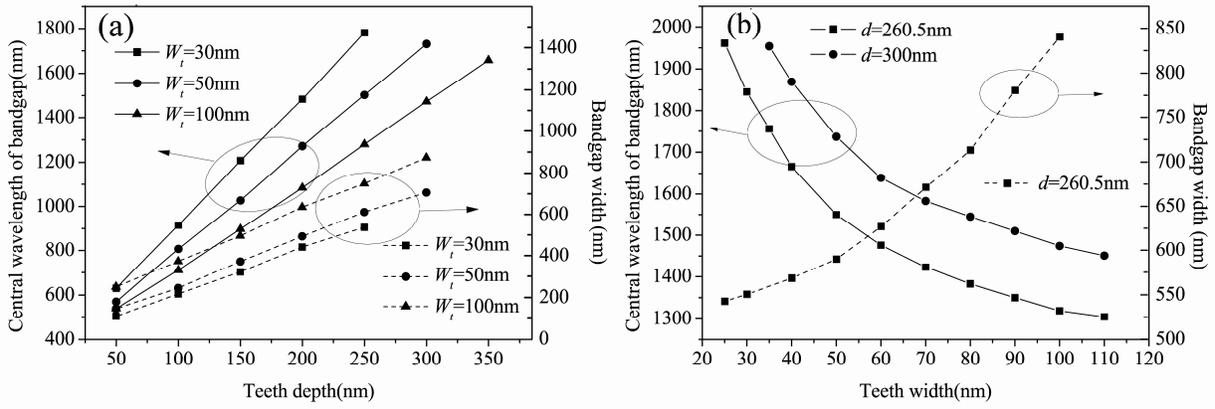

**Fig. 7**

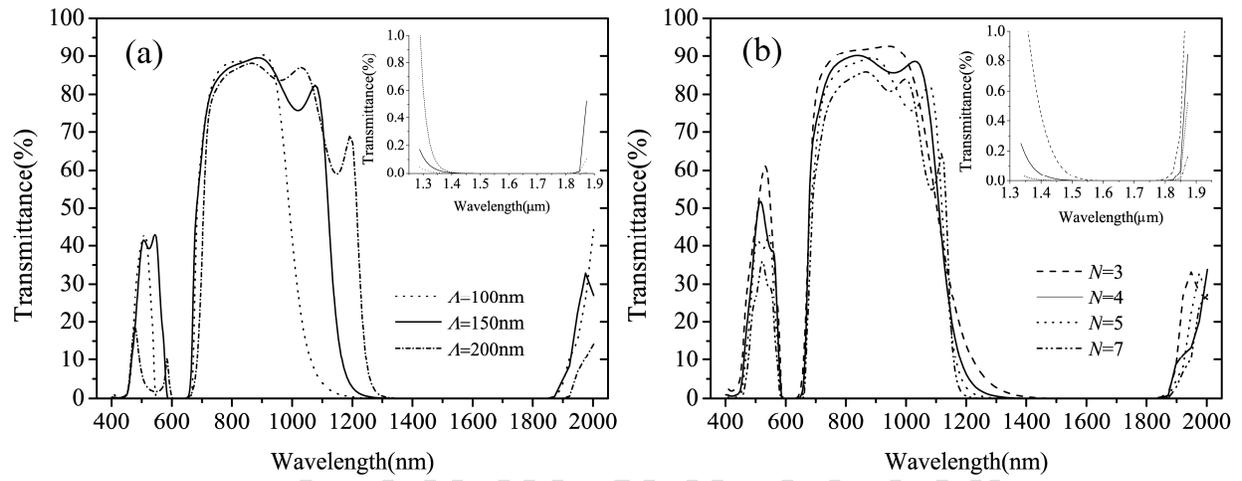